\documentclass[a4paper]{article}
\usepackage[utf8]{inputenc}
\usepackage[T1]{fontenc}
\usepackage{fixltx2e}
\usepackage{graphicx}
\usepackage{longtable}
\usepackage{float}
\usepackage{wrapfig}
\usepackage{rotating}
\usepackage[normalem]{ulem}
\usepackage{amsmath}
\usepackage{textcomp}
\usepackage{marvosym}
\usepackage{wasysym}
\usepackage{amssymb}
\usepackage{hyperref}
\author{A. Feigel \\ Racah Institute of Physics, Hebrew University of Jerusalem \\ sasha@phys.huji.ac.il}
\date{\today}
\title{Dynamics of a Mechanical System with Multiple Degrees of Freedom out of Thermal Equilibrium.}
\hypersetup{
  pdfkeywords={},
  pdfsubject={},
  pdfcreator={Emacs 24.5.1 (Org mode 8.2.10)}}
\begin{document}

\maketitle
\begin{abstract}
Out of thermal equilibrium, an environment imposes effective mechanical forces on microscopical nanofabricated devices, chemical or biological systems. Here we address the question of how to calculate these forces together with the response of the system from the first principles. We show that an ideal gas-like environment, even near thermal equilibrium, can enforce a specific steady state on the system by creating effective potentials in otherwise homogeneous space. An example of stable and unstable rectifications of thermal fluctuations is presented using modified Feynman-Smoluchowski ratchet with two degrees of freedom. Moreover, the stability of a steady configuration depends on its chiral symmetry. The transition rate probabilities and the corresponding kinetic equations are derived for a complex mechanical system with arbitrary degrees of freedom. This work therefore extends the applicability of mechanical systems as a toy model playground of statistical physics for multiple degrees of freedom active and living matter.
\end{abstract}

Macroscopic machines become unstable and disintegrate after some threshold load. Stability is, therefore, an important factor in their design. The same question will be important for nanodevices\cite{Haenggi2009} when they will develop to maximum capacity. Stability of the future complex devices is a fundamental question of statistical physics regarding nonlinear dynamics out of thermal equilibrium. Dynamics depends on the interaction between the system's degrees of freedom and the forces that drive it out of thermal equilibrium.

A non-equilibrium environment affects degrees of freedom of a small mechanical system. A chemical or nanofabricated device out of thermal equilibrium can experience effective forces even in the absence of external potentials\cite{Seifert2012}. There exist Brownian machines with a single relevant degree of freedom, such as Feynman-Smoluchowski ratchet\cite{Smoluchowski1912,Feynman1966}, capable of utilizing these effective forces to generate useful work out of thermal fluctuations. There is a significant knowledge gap as to how the same forces affect dynamics of a system with multiple degrees of freedom, though there are theoretical and experimental efforts in this direction\cite{Serra-Garcia2016,Ryabov2016}.

The subject of this work is a mechanical system composed of solid bodies which are connected by rigid axis or free rotating joints\cite{VandenBroeck2004,Lipson2000}. The different parts of the system, preserving the connectivity, are immersed in the reservoirs containing ideal gases. Elastic collisions with the gas particles cause the transitions of the system between its states. In the absence of external potentials, any configuration of a mechanical system possesses the same probability at thermal equilibrium under the condition of thermodynamic limit, meaning the temperatures of all bathes are equal, their volumes are large and mass of a gas particle is negligible relative to the mass of the system. Let us address the question whether specific configuration may be favored out of thermal equilibrium and how to calculate this configuration as a function of the properties of the system?

An advantage of a mechanical framework for a thermodynamic system is the possibility for the treatment from the first principles by modeling the thermal bath as a gas of small elastic particles\cite{VandenBroeck2004,Zheng2010}. The disadvantage is the apparent difference of mechanical framework from the real and artificial nanosystems that are driven by chemical or optical sources out of thermal equilibrium\cite{Haenggi2009}. Nevertheless, mechanical Feynman-Smoluchowski ratchet was declared as an inspiration for some chemical\cite{Hernandez2004,Erbas-Cakmak2015} and mechanical devices\cite{Blickle2012,Martinez2016,Serra-Garcia2016}. Ratchet mechanism is considered  important for verification of fundamental theorems of statistical physics\cite{Parrondo1996,Sekimoto1997,Magnasco1998} and an essential property of molecular motors\cite{Bier1997,Astumian1997,Bar-Nahum2005,Brueckner2008,Spirin2009,Sousa2005,Dangkulwanich2013a} (though alternative hypotheses regarding motion of molecular motors do exist).

To the best of our knowledge, only the systems with a single degree of freedom were considered from the first principles so far\cite{VandenBroeck2004,Broek2008}. A system with at least two degrees of freedom, however, is required to investigate the phenomenon of interacting degrees of freedom and influence of this phenomenon on dynamics of a complex system out of thermal equilibrium. 

In this Article, we present ab initio path from the elastic scattering of a single gas particle by a mechanical system, to transition rate probability between the states of the system, to the corresponding Masters Boltzmann equation and average velocities of the system's degrees of freedom as functions of macroscopic parameters of the out of equilibrium environment (Onsager relations)\cite{Onsager1931} including influence of the different degrees of freedom on each other. Stability of the steady state of the system depends on the interaction between its degrees of freedom and the effective forces imposed on the system by the environment. An interesting finding is that some of these forces persist even in a single temperature environment if the thermodynamic limit does not hold. In addition, spatial asymmetry of the system's stable state, together with the corresponding directed motion, may possess preferred chiral symmetry. To make the discussion more visual we demonstrate all these phenomena using a modified Feynman-Smoluchowski ratchet with two degrees of freedom.

\begin{figure}[htb]
\label{fig1}
\centering
\includegraphics[width=.9\linewidth]{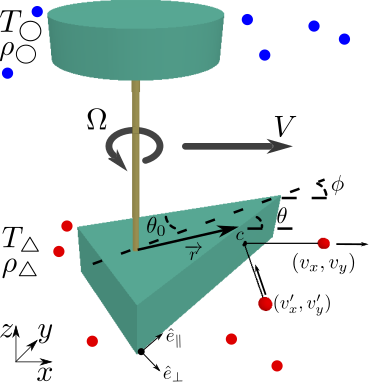}
\caption{Dumbbell structure with the asymmetric part in the form of an isosceles triangle with apex angle $2\theta_{0}$, which is connected to a symmetric part by rigid axis. There are two degrees of freedom translational velocity $V=\partial x /\partial t$, along $x$ coordinate and frequency $\Omega=\partial\phi /\partial t$, where $\phi$ is the angle between the median of the triangle and axis $x$. The symmetric part is immersed in a thermal bath composed of an ideal gas with elastic particles (blue circles) of mass $m$ at temperature $T_{\bigcirc}$ and density $\rho_{\bigcirc}$. The symmetric part is in interaction with the particles  (red points) of mass $m$ at temperature $T_{\bigtriangleup}$ and density $\rho_{\bigtriangleup}$. Dynamics of the dumbbell structure is driven by scattering of the gas particles. Scattering from velocity $(v'_{x},v'_{y})$ to $(v_{x},v_{y})$ may occur at any point $c$ along the surface of the structure. Change in the velocity and rotation frequency of the structure due to a single scattering event depends on the geometry of the surface of the structure at the point of scattering $c$, such as an angle $\theta$ between axis $x$ and tangential of the surface and vector $\protect\overrightarrow{r}$ from axis of rotation to the point $c$. The total change in momentum and in the energy of the structure depends on cumulative effect of the interactions at all points $c$ along its surface. The structure is a Brownian motor and may possess average velocity $<V>\neq 0$ which takes its maximum absolute values when the triangle is directed along axis $x$, $\phi=0,\pi$. Velocity $<V>$ vanishes due to symmetry at $\phi=\pi/2,-\pi/2$. The main question is: What is the stable direction $\phi$ of the dumbbell structure as a function of the thermal bath temperatures, the shape of the asymmetric part $\theta(c)$ and position of rotation axis?}
\end{figure}

Consider a dumbbell-like macroscopic body consisting of symmetric and asymmetric  parts, see \ref{fig1}. These parts are rigidly connected to each other by a thin axis. Each part is in contact with a dedicated thermal bath which constitutes an infinite reservoir of the ideal gas composed of identical particles of mass $m$ at density $\rho_{\bigtriangleup}$ and temperature $T_{\bigtriangleup}$ around the asymmetric part and density $\rho_{\bigcirc}$ and temperature $T_{\bigcirc}$ around symmetric part. Gas particles, in each thermal bath, move only in $(x,y)$ plane with velocities $(v_{x},v_{y})$ at Maxwell distribution with temperatures $T_{\bigtriangleup}$ and $T_{\bigcirc}$ correspondingly. The body possesses two degrees of freedom: First, it can move along $x$ axis with velocity $V$. Second, it can rotate around $z$ axis with frequency $\Omega$. The rotation angle coordinate is $\phi$. The kinetic 
energy of the body, therefore, is $MV^{2}/2+I\Omega^{2}/2$, where $M$ is the mass and $I$ is the moment of inertia. Both degrees of freedom are translation invariant because no external potential is present. The system is a multiple degree of freedom version of a single degree of freedom Triangulita motor\cite{VandenBroeck2004}, which is an elegant and simple version of the Feynman-Smoluchowski ratchet. 

To analyze dynamics and stability of dumbbell structures, one should calculate velocity $V$ and frequency $\Omega$ as a function of the corresponding coordinates $x$ and $\phi$. The first task is to derive transition probabilities $W(V,V')$ and $W(\Omega,\Omega')$ from velocity $V'\rightarrow V$ and frequency $\Omega'\rightarrow \Omega$ due to elastic scattering of the gas particles\cite{VandenBroeck2004,Broek2008,Meurs2004}. A single particle may scatter at any point $c$ along the perimeter of the motor. Then, transition probability rates $W$ are averaged over all interaction in all points $c$ and all velocities of the gas particles. Finally, the average moment of $<V^{n}>$ and $<\Omega^{n}>$ as the functions of $x$ and $\phi$ are calculated from $W$ using Kramers-Moyal expansion. 

Consider the scattering of a single gas particle with mass $m$ by a macroscopic body with mass $M$ and moment of inertia $I$. In the frame of reference of the body, conservation of energy, momentum, angular momentum, and tangential velocity of the particle along the surface of the body are:
\begin{eqnarray}
  \label{eq:rotsys12}
  &&-MV^2-I\Omega^2+m(v'_x-v_x)(v'_x+v_x)+\\\nonumber
  &&m(v'_y-v_y)(v'_y+v_y)=0,\\\nonumber
  &&m(v'_x-v_x)=MV,\\\nonumber
  &&I\Omega+|r\times v|=|r\times v'|,\\\nonumber
  &&v'_x\sin\theta+v'_y\cos\theta=v_x\sin\theta+v_y\cos\theta,
\end{eqnarray}
where $(v'_{x},v'_{y})$ are the velocities of the particle prior collision. After the collision, velocities of the gas particle together with the velocity and the rotational frequency of the body are $(v_{x},v_{y})$, $V$ and $\Omega$ correspondingly. Angle $\theta$ and radius $\overrightarrow{r}$ depend on the body's geometry and point of the scattering along the surface of the body, see Figure \ref{fig1}.

Let us define two geometric factors:
\begin{eqnarray}
\label{eq:33}
 \Gamma_{V}(c) = \sin\theta,\;\;\Gamma_{\Omega}(c) = r_{x}\cos\theta+r_{y}\sin\theta,
\end{eqnarray} 
where $c$ indicates a point of the scattering on the body's surface and, therefore, defines angle $\theta$ and radius $\overrightarrow{r}=(r_{x},r_{y})$.  This is a natural choice because the factors (\ref{eq:33}) define linear viscous coefficients for translation and rotation degrees of freedom correspondingly.  On an analogy level only, connection of $\Gamma$ to the linear viscous coefficients may be inferred from the similarity between the equalities:
\begin{eqnarray}
  \label{eq:407876} m^{2}\Delta v_{x}^{2}+m^{2}\Delta v_{y}^{2}=M^{2}V^{2}\frac{1}{\Gamma_{V}^{2}(c)}=I^{2}\Omega^{2}\frac{1}{\Gamma_{\Omega}^{2}(c)},
\end{eqnarray}
where $\Delta v_{i}=v_{i}-v'_{i}$, and the expression for diffusion coefficient of the body in velocity space according to the fluctuation dissipation theorem\cite{Kubo1966}:
\begin{eqnarray}
  \label{eq:44}
  D_{V}=\left <\frac{V^{2}}{\Delta t}\right >=\frac{T\gamma_{V}}{M},
\end{eqnarray}
where $D_{V}$ is the diffusion coefficient in the velocity space, $\gamma_{V}$ is the linear viscous coefficient. The similarity of (\ref{eq:407876}) and (\ref{eq:44}) is evident under the assumption that $\sum \Delta v_{i}^{2}\propto T$ at thermal equilibrium. Equalities (\ref{eq:407876}) follow from (\ref{eq:rotsys12}). We will see later that these equalities and their connection to the linear viscous coefficients hold rigorously in the case of the arbitrary number of degrees of freedom.

After the transition to the lab frame of references by adding velocity $(-V'+\Omega' r_{y},-\Omega' r_{x})$ and rotational frequency $-\Omega$ to the degrees of freedom of the system, one gets:
\begin{eqnarray}
  \label{eq:42uuy} -\Delta V-\frac{M}{m}\Delta V\frac{G^{2}_{V}(c)}{\Gamma^{2}_{V}(c)}-2V'+\Omega'\frac{\Gamma_{\Omega}}{\Gamma_{V}} +2v'_x-2v'_y\frac{1}{\tan\theta}=0,
\end{eqnarray}
where $\Delta V=V-V'$ and:
\begin{eqnarray}
\label{eq:37768} -\Delta\Omega-\frac{I}{m}\Delta\Omega\frac{G^{2}_{\Omega}(c)}{\Gamma^{2}_{\Omega}(c)}-2\Omega'+V'\frac{\Gamma_{V}}{\Gamma_{\Omega}}-2v'_{x}\frac{1}{r_{y}+r_{x}/\tan{\theta}}+2v'_{y}\frac{1}{r_{y}\tan{\theta}+r_{x}}=0,
\end{eqnarray}
where $\Delta\Omega=\Omega-\Omega'$ and the coefficients $G$ are:
\begin{eqnarray}
  \label{eq:53879}  G^{2}_{V}=1+\frac{m}{I}\Gamma^{2}_{\Omega},\;G^{2}_{\Omega}=1+\frac{m}{M}\Gamma^{2}_{V},
\end{eqnarray}
describe interaction between degrees of freedom imposed by the environment. The terms $\Omega'\Gamma_{\Omega}/\Gamma_{V}$ and $V'\Gamma_{V}/\Gamma_{\Omega}$ correspond to the direct interaction of the degrees of freedom.

Eqs. (\ref{eq:42uuy}) and (\ref{eq:37768}) can be written in a unified form for the case of an arbitrary number of degrees of freedom. For each degree of freedom $\xi$ holds:
\begin{eqnarray}
  \label{eq:kihkjhk}  
-\Delta\dot{X}_{\xi}\left (1+\frac{M_{\xi}}{m}\left (\frac{G_{\xi}(c)}{\Gamma_{\xi}(c)}\right )^{2}\right )-2\dot{X}'_{\xi}+\sum_{\xi'\neq\xi}\frac{\Gamma_{\xi'}(c)}{\Gamma_{\xi}(c)}\dot{X}'_{\xi'}+g_{x,\xi}(c)v'_x+g_{y,\xi}(c)v'_y=0,
\end{eqnarray}
where $\dot{X}_{\xi}$ is the velocity such as $V$ or $\Omega$, $M_{\xi}$ is the mass such is $M$ or $I$. Geometric factors $\Gamma$ depend on interaction channel with the thermal bath $c$ such as point of collision in the case of the dumbbell structure. The affect of other degrees of freedom comes in the resale of the mass $M_{\xi}$ by the factor:
\begin{eqnarray}
  \label{eq:597965}
  G^{2}_{\xi,i}(c)=1+\sum_{\xi'\neq \xi}\frac{m}{M_{\xi',i}}\Gamma^{2}_{\xi',i}(c),
\end{eqnarray}
and by the update of the velocity $\dot{X}'_{\xi}$ by other degrees of freedom $\sum_{\xi'\neq\xi}\Gamma_{\xi'}/\Gamma_{\xi}\dot{X}'_{\xi'}$. The velocities $v'_{x}$ and $v'_{y}$ are velocities of the gas particle before the collision. The weights $g_{x}$ and $g_{y}$ fit the condition:
\begin{eqnarray}
\label{eq:11}
g^{2}_{x,\xi}+g^{2}_{y,\xi}=\frac{1}{\Gamma^{2}_{\xi}}
\end{eqnarray} 
This condition holds both for (\ref{eq:42uuy}) and (\ref{eq:37768}). Later we will see that it is connected to the detailed balance at thermal equilibrium. 

The averaging of (\ref{eq:kihkjhk}) over all possible velocities of the colliding gas particles results in transition rate probability $W$:
\begin{eqnarray}
  \label{eq:7589}
  &&W(\dot{X}_{\xi},\Delta \dot{X}_{\xi})=\frac{1}{4}\sum_{i}\rho_{i}\sqrt{\frac{m}{2\pi T_{i}}}\oint dc_{i}\\\nonumber
&&\left | \Delta \dot{X}_{\xi}\right |H\left [\Delta\dot{X}_{\xi}\Gamma_{\xi,i}\left (1+\frac{M_{\xi}}{m}\left (\frac{G_{\xi,i}(c)}{\Gamma_{\xi,i}(c)}\right )^{2}\right )-\sum_{\xi'\neq\xi}\dot{X}_{\xi'}\Gamma_{\xi',i}(c)\right ]\\\nonumber
&&\Gamma^{2}_{\xi,i}(c)\left (1+\frac{M_{\xi}}{m}\left (\frac{G_{\xi}(c)}{\Gamma_{\xi,i}(c)}\right )^{2}\right )^{2}\\\nonumber
&&\exp\left [-\frac{m\Gamma^{2}_{\xi,i}(c)\left ( \dot{X}_{\xi}-\frac{1}{2}\sum_{\xi'\neq \xi}\frac{\Gamma_{\xi',i}}{\Gamma_{\xi,i}}\dot{X}_{\xi'}+\frac{1}{2}\left[ \Delta \dot{X}_{\xi}\left (\frac{M_{\xi} G^{2}_{\xi,i}(c)}{m\Gamma_{\xi,i}^{2}(c)}+1\right )\right ]\right )^{2}}{2T_{i}}\right ]
\end{eqnarray}
where index $i$ goes over all thermal bathes; see Appendix for  details. In the limit $G=1$ and $\sum_{\xi'\neq\xi}=0$ expression (\ref{eq:7589}) converges to the results obtained for a single degree of freedom systems\cite{VandenBroeck2004,Broek2008}.

Transition rate probability in the form of (\ref{eq:7589}) makes it possible to calculate average velocities  $<\dot{X}^{n}>$ of a mechanical system as a function of velocities of other degrees of freedom, external forces and forces as a consequence of out of equilibrium environment. It is done using Kramers-Moyal expansion of the corresponding Masters Boltzmann equation for the probability to possess specific velocity:
\begin{eqnarray}
  \label{eq:34}
  \frac{\partial P(\dot{X},t)}{\partial t} = &&\int W(\dot{X}-\Delta \dot{X},\Delta \dot{X})P(\dot{X}-\Delta \dot{X},t)d\Delta \dot{X} - \\\nonumber
&&P(\dot{X},t)\int W(\dot{X},-\Delta \dot{X})d\Delta \dot{X},
\end{eqnarray}
where probability $P(\dot{X},t)$ for velocity $\dot{X}$ at time $t$. This description is valid in the overdumped regime and if the velocities are uncorrelated.  Therefore, we omit index $\xi$.

The first three moments are:
\begin{eqnarray}
  \label{eq:357657}
  \frac{\partial <\dot{X}>}{\partial t}&=&<a_{1}(\dot{X})>,\\\nonumber
  \frac{\partial <\dot{X}^{2}>}{\partial t}&=&2<\dot{X}a_{1}(\dot{X})>+<a_{2}(\dot{X})>,\\\nonumber
  \frac{\partial <\dot{X}^{3}>}{\partial t}&=&3<\dot{X}^{2}a_{1}(\dot{X})>+3<\dot{X}a_{2}(\dot{X})>+<a_{3}(\dot{X})>,
\end{eqnarray}
where coefficients $a$ defined by Kramers-Moyal expansion:
\begin{eqnarray}
  \label{eq:34}
  \frac{\partial P(\dot{X},t)}{\partial t} = \sum_{n=1}^{\infty}\frac{(-1)^{n}}{n!}\frac{d^{n}}{d\dot{X}^{n}}\left [a_{n}(\dot{X})P(\dot{X},t) \right ],
\end{eqnarray}
where:
\begin{eqnarray}
\label{eq:12}
 a_{n}(\dot{X}) = \int \Delta \dot{X}^{n}W(\dot{X},\Delta \dot{X})d\Delta \dot{X},
\end{eqnarray}
where $W$ defined by (\ref{eq:309837}).

To demonstrate the influence of thermal fluctuations in the system and to simplify the presentation, let us consider Kramers-Moyal expansion (\ref{eq:357657}) in the case $\dot{X}_{\xi'}=0$ for $\xi'\neq\xi$. Velocities of other degrees of freedom impose drag force that can be added later. The first three moments of the Kramers-Moyal expansion in this case are:
\begin{eqnarray}
  \label{eq:52}
&&\frac{\partial <\dot{X}>}{\partial t} = \sum_{i}\rho_{i}\sqrt{\frac{T_{i}}{m}}\\\nonumber
&&\left [ -\sqrt{\frac{T_{i}}{M_{X}}}\oint\frac{\Gamma_{\dot{X},i}(c)}{G^{2}_{\dot{X},i}(c)}\epsilon_{\dot{X}}^{1}\right .\\\nonumber
&&-2\sqrt{\frac{2}{\pi}}\oint \frac{\Gamma_{\dot{X},i}^{2}(c)}{G_{\dot{X},i}^{2}(c)}<\dot{X}>\epsilon_{\dot{X}}^{2}\\\nonumber
&&\left . +\left ( \oint\frac{\Gamma_{\dot{X},i}^{3}}{G_{\dot{X},i}^{4}}\sqrt{\frac{T_{i}}{M_{X}}}-\oint\frac{\Gamma_{\dot{X},i}^{3}}{G_{\dot{X},i}^{2}}\sqrt{\frac{M_{X}}{T_{i}}}<\dot{X}^{2}>\right ) \epsilon_{\dot{X}}^{3} \right ],
\end{eqnarray}
the second order:
\begin{eqnarray}
  \label{eq:524534}
&&\frac{\partial <\dot{X}^{2}>}{\partial t} = \\\nonumber
&&\sum_{i} \rho_{i}\sqrt{\frac{T_{i}}{m}} \left [ -2\sqrt{\frac{T_{i}}{M_{X}}}\oint\frac{\Gamma_{\dot{X},i}}{G^{2}_{\dot{X},i}}<\dot{X}>\epsilon_{\dot{X}}^{1}\right .\\\nonumber
&&\left .+4\sqrt{\frac{2}{\pi}}\left( \oint\frac{\Gamma_{\dot{X},i}^{2}}{G_{\dot{X},i}^{4}}\frac{T_{i}}{M_{X}}-\oint\frac{\Gamma_{\dot{X},i}^{2}}{G^{2}_{\dot{X},i}}<\dot{X}^{2}>\right )\epsilon_{\dot{X}}^{2} \right .\\\nonumber
&&\left . -2\left(-4 \oint\frac{\Gamma_{\dot{X},i}^{3}}{G_{\dot{X},i}^{4}}\sqrt{\frac{T_{i}}{M_{X}}}<\dot{X}>+\sqrt{\frac{M_{X}}{T_{i}}}\oint\frac{\Gamma_{\dot{X},i}^{3}}{G^{2}_{\dot{X},i}}<\dot{X}^{3}>\right )\epsilon_{\dot{X}}^{3} \right ],
\end{eqnarray}
the third order:
\begin{eqnarray}
  \label{eq:52479}
&&\frac{\partial <\dot{X}^{3}>}{\partial t} = \\\nonumber
&&\sum_{i} \rho_{i}\sqrt{\frac{T_{i}}{m}} \left [ -3\sqrt{\frac{T_{i}}{M_{X}}}\oint\frac{\Gamma_{\dot{X},i}}{G^{2}_{\dot{X},i}}<\dot{X}^{2}>\epsilon_{\dot{X}}^{1}\right .\\\nonumber
&&\left .+6\sqrt{\frac{2}{\pi}}\left( 2\oint\frac{\Gamma_{\dot{X},i}^{2}}{G_{\dot{X},i}^{4}}\frac{T_{i}}{\sqrt{M_{X}}}<\dot{X}>-\sqrt{M_{X}}\oint\frac{\Gamma_{\dot{X},i}^{2}}{G^{2}_{\dot{X},i}}<\dot{X}^{3}>\right )\epsilon_{\dot{X}}^{2} \right ],
\end{eqnarray}
where $\epsilon_{\dot{X}}=m/M_{X}$, index $i$ goes over the thermal bathes and the index $\xi$ is omitted. The expressions (\ref{eq:52}),(\ref{eq:524534}) and (\ref{eq:52479}) are derived with the help of Wolfram Mathematica software.

If there is no affect of the other degrees of freedom $G=1$, eqs. (\ref{eq:52}),(\ref{eq:524534}) and (\ref{eq:52479}) converge to the corresponding results of a system with single degree of freedom\cite{Meurs2004}, taking into account that $\oint\Gamma=0$. The non vanishing term proportional to $\Gamma/G^{2}$ is a unique property of multiple degrees of freedom systems. 

Leading term for the average velocity at a steady state follows from (\ref{eq:52}),(\ref{eq:524534}) and (\ref{eq:52479}), after neglecting the time derivatives, as:
\begin{eqnarray}
  \label{eq:1}
<\dot{X}_{\xi}>_{\Gamma} = -\frac{1}{2\sqrt{M_{\xi}}}\sqrt{\frac{\pi}{2}}\frac{\sum_{i}\rho_{i}T_{i}\oint\frac{\Gamma_{\dot{X}_{\xi},i}(c)}{G^{2}_{\dot{X}_{\xi},i}(c)}}{\sum_{i}\rho_{i}T^{\frac{1}{2}}_{i}\oint \frac{\Gamma_{\dot{X}_{\xi},i}^{2}(c)}{G_{\dot{X}_{\xi},i}^{2}(c)}}\epsilon_{\dot{X}_{\xi}}^{-1}.  
\end{eqnarray}
It is of the order $\epsilon_{\dot{X}_{\xi'}}^{2}/\epsilon_{\dot{X}_{\xi}}$ taking into account that $\oint\Gamma=0$ and $1/G_{\dot{X}_{\xi}}^{2}\approx 1-\epsilon_{\dot{X}_{\xi'}}^{2}\Gamma_{\dot{X}_{\xi'}}^{2}$. This results vanishes when  $G=1$, e.g. in the systems with a single degree of freedom.

The next order velocity correction is:
\begin{eqnarray}
  \label{eq:2}
&&<\dot{X}_{\xi}>_{\Gamma^{3}} =\\\nonumber
&&\frac{1}{2}\sqrt{\frac{\pi}{2}}\frac{ \sum_{i}\rho_{i}\left ( T_{i}\oint\frac{\Gamma_{\dot{X}_{\xi},i}^{3}}{G_{\dot{X}_{\xi},i}^{4}}-M_{\xi}\dot{X}_{0\xi}^{2}\oint\frac{\Gamma_{\dot{X}_{\xi},i}^{3}}{G_{\dot{X}_{\xi},i}^{2}}\right )}{ \sum_{i}\rho_{i}T^{\frac{1}{2}}_{i}\oint \frac{\Gamma_{\dot{X}_{\xi},i}^{2}(c)}{G_{\dot{X}_{\xi},i}^{2}(c)}
 }\frac{\epsilon_{\dot{X}_{\xi}}}{\sqrt{M_{\xi}}},
\end{eqnarray}
where:
\begin{eqnarray}
  \label{eq:16} \dot{X}^{2}_{0\xi}=\frac{1}{M_{\xi}}\frac{\sum_{i}\rho_{i}T_{i}^{\frac{3}{2}}\frac{\Gamma_{\dot{X}_{\xi},i}^{2}}{G_{\dot{X}_{\xi},i}^{4}}}{\sum_{i}\rho_{i}T_{i}^{\frac{1}{2}}\frac{\Gamma_{\dot{X}_{\xi},i}^{2}}{G_{\dot{X}_{\xi},i}^{2}}}.
\end{eqnarray}
This term describes rectified Brownian velocity and remains finite even in the case of a single degree of freedom\cite{VandenBroeck2004,Broek2008}.

If the temperatures of all thermal bathes are equal $T_{i} = T_{j}$, Brownian velocity (\ref{eq:2}) vanishes, while  velocity (\ref{eq:1}) remain finite. It is still an out of equilibrium phenomenon because (\ref{eq:1}) vanishes at the thermodynamics limit $m/M\rightarrow 0$.

One gets the drag force imposed on degree of freedom $\xi$ by all other degrees of freedom $\xi'\neq\xi$ introducing finite velocities of other degrees of freedom $\dot{X}_{\xi}\rightarrow\dot{X}_{\xi}-\frac{1}{2}\sum_{\xi'\neq \xi}\frac{\Gamma_{\xi',i}}{\Gamma_{\xi,i}}\dot{X}_{\xi'}$ to (\ref{eq:52}) : 
\begin{eqnarray}
  \label{eq:1453}
<\dot{X}_{\xi}> =\frac{1}{2} \frac{\sum_{\xi'\neq\xi}\sum_{i}\rho_{i}T_{i}<\dot{X}_{\xi'}>\oint\frac{\Gamma_{\dot{X}_{\xi},i}(c)\Gamma_{\dot{X}_{\xi'},i}(c)}{G^{2}_{\dot{X}_{\xi},i}(c)}}{\sum_{i}\rho_{i}T^{\frac{1}{2}}_{i}\oint \frac{\Gamma_{\dot{X}_{\xi},i}^{2}(c)}{G_{\dot{X}_{\xi},i}^{2}(c)}}. 
\end{eqnarray}
It is important to note that substituting (\ref{eq:3785554}) in (\ref{eq:1453}) one get the same order of magnitude as (\ref{eq:1}).

Following previous general results, the dynamics of a dumbbell structure near the point $\phi=0$ can be presented as Onsager relations with non-linear corrections due to interaction between translational and rotational degrees of freedom:
\begin{eqnarray}
  \label{eq:ytuyt456}
  \frac{\partial x}{\partial t} &=& \frac{F_{x}}{\gamma_{x}}+<V>_{rec}+L_{x,\phi}\frac{\partial \phi}{\partial t},\\\nonumber
  \frac{\partial \phi}{\partial t} &=& \frac{M}{\gamma_{\phi}}+L_{\phi,\phi}\phi+\hat{L}_{\phi,x}\phi\frac{\partial x}{\partial t},   
\end{eqnarray}
where $L_{x,\Delta T}$ and $L_{\phi,\phi}$ are Onsager coefficients, $\gamma$ is linear viscosity, $<V>$ is rectified velocity (\ref{eq:2}), $L_{\phi,xn}=\hat{L}_{\phi,x}\phi$ and $L_{x,\phi}$ mutual influence of degrees of freedom (\ref{eq:1453}). The Onsager coefficient $L_{\phi,\phi}$  corresponds to effective force (\ref{eq:1}) that vanishes at $\phi=0$ due to symmetry.  

Linear dissipation coefficient $\gamma_{\xi}=\sum_{i}\gamma_{\xi,i}$ follows from (\ref{eq:52}):
\begin{eqnarray}
  \label{eq:337587} \gamma_{\xi}=\sum_{i}4\rho_{i}\sqrt{\frac{mT_{i}}{2\pi}}\oint_{c}\Gamma_{\dot{X}_{\xi},i}^{2}(c).
\end{eqnarray}
At this moment, one can see the connection between $\Gamma^{2}$ and the linear dissipation coefficient.

Transnational velocity of dumbbell structure following (\ref{eq:2}) is:
\begin{eqnarray}
  \label{eq:3785554}
  <V>_{rec} = \sqrt{\frac{m}{M}}\sqrt{\frac{\pi }{8M}}\frac{\sum_{i}\rho_{i}(T_{i}-MV^{2}_{0})\oint_{c}\Gamma^{3}_{V,i}(c)}{\sum_{i}\rho_{i}T^{1/2}_{i}\oint_{c}\Gamma_{V,i}^{2}(c)},.
\end{eqnarray}
Near thermal equilibrium, it can be presented as a function of the affinity $\Delta T/T^{2}$:
\begin{eqnarray}
\label{eq:4}
 <V>_{rec}=L_{x,\phi}\frac{\Delta T}{T^{2}}.
\end{eqnarray}
It corresponds to the result\cite{VandenBroeck2004,Broek2008,Meurs2004}. Maximum velocity is achieved when $\phi=0$ or $\phi=\pi$. No motion occurs for $\phi=\pi/2$ and $\phi=-\pi/2$ because $\oint\Gamma^{3}_{V}$ vanishes in these cases.

Rotation dynamics of the dumbbell structure depends on the interaction with translation degree of freedom (\ref{eq:1453}):
\begin{eqnarray}
\label{eq:14}   L_{\phi,x}=\frac{1}{2}\frac{S_{\bigtriangleup}\rho_{\bigtriangleup} T_{\bigtriangleup}^{1/2}\oint\Gamma_{\Omega,\bigtriangleup}\Gamma_{V,\bigtriangleup}}{\sum_{i=\bigtriangleup,\bigcirc}\rho_{i} T_{i}^{1/2}\oint\Gamma^{2}_{\Omega,i}},
\end{eqnarray}
and the effective forces imposed by the out-of-equilibrium environment (\ref{eq:1}):
\begin{eqnarray}
  \label{eq:22}  <\Omega>_{rec}=\frac{1}{2}\frac{\sqrt{m}}{M}\frac{S_{\bigtriangleup}\rho_{\bigtriangleup} T_{\bigtriangleup}^{1/2}\oint\Gamma_{\Omega,\bigtriangleup}\Gamma^{2}_{V,\bigtriangleup}}{\sum_{i=\bigtriangleup,\bigcirc}\rho_{i} T_{i}^{1/2}\oint\Gamma^{2}_{\Omega,i}},
\end{eqnarray}
which can be written as:
\begin{eqnarray}
\label{eq:5}
 <\Omega>_{rec}=L_{\phi,\phi}\phi,
\end{eqnarray}
near $\phi=0$.

Following (\ref{eq:ytuyt456}), in the absence of external force, the condition for stable direction $\phi=0$ of the dumbbell structure is:
\begin{eqnarray}
\label{eq:6}
 L_{\phi,\phi}+\hat{L}_{\phi,x}\frac{\partial x}{\partial t}<0.
\end{eqnarray}
The state of the system changes from stable to unstable and vice versa at critical velocity $V_{cr}=\partial x/\partial t$:
\begin{eqnarray}
\label{eq:3}
V_{cr}=-\frac{L_{\phi,\phi}}{\hat{L}_{\phi,x}}=-\lim_{\phi\rightarrow 0}\frac{\sqrt{m}}{M}\frac{\oint\Gamma_{\Omega}\Gamma_{V}^{2}}{\oint\Gamma_{\Omega}\Gamma_{V}}.
\end{eqnarray}
The dumbbell structure may possess transition (\ref{eq:3}).

Let us first consider the dynamic of dumbbell structure with $\theta_{0}=\pi/24$ and axis of rotation near the apex, see Figure 2. This figure presents velocity (\ref{eq:3785554})and stability (\ref{eq:6}) as the functions of temperature difference $T_{\bigcirc}-T_{\bigtriangleup}$, under constraint that $T_{\bigcirc}+T_{\bigtriangleup}=20$. The velocity of the triangle with a median along the $x$ axis $\phi=0,\pi$ is directed toward the apex if $T_{\bigtriangleup}<T_{\bigcirc}$ and toward the base otherwise $T_{\bigtriangleup}>T_{\bigcirc}$. Motion towards the base become unstable beyond critical velocity $V_{cr}$ (\ref{eq:3}). In this case, stable position of the triangle's median is perpendicular to the axis $x$, $\phi=\pm\pi/2$. Motion in the direction of the apex is always stable. 

Stability depends on the gases' temperatures, shape of the structure, and position of the rotation axis. Consider two cases when the axis of rotation is located near the apex and near the base on the median of the triangle. Figure \ref{fig3}, presents stability in  both cases as a function of apex angle $\theta_{0}$ together with temperature difference  $T_{\bigcirc}-T_{\bigtriangleup}$. If rotation axis located near the apex of the triangle, then the motion in either direction is mainly stable, but for a small region of high velocities toward the base, as in Fig. 2. At the same region of high velocity in the direction of the base, however, the motion is stable if the axis of rotation located near the base.In addition, In the case of rotation axis near the base,  
the motion is unstable if apex angle is acute $2\theta_{0}<\pi/2$ and is stable for obtuse triangle $2\theta_{0}>\pi/2$.

Dynamics of dumbbell structure possesses intuitive and counter-intuitive features. Intuitively at the macroscopic level, the motion of a triangle with the axis of rotation near its apex is stable, while the motion of the same triangle in the direction of its base is unstable. In this case, motion toward the base creates momentum (\ref{eq:14}) that moves the triangle from this direction. Counter-intuitive to macroscopic intuition motion of the triangle with the axis of rotation near its base toward the corner can be stable. The motion becomes stable due to effective potential (\ref{eq:22}) imposed on the system by the out of thermal equilibrium environment. This motion becomes unstable only above some critical velocity when imposed momentum becomes high. This instability occurs only if the apex angle is acute enough to achieve this velocity.

The main finding of this work is the important role of local potentials imposed on the system by out of thermal equilibrium environment. The potential imposed by the out of thermal equilibrium environment on a mechanical system corresponds to  the Onsager coefficient, such as $L_{\phi,\phi}$ in (\ref{eq:ytuyt456}). Moreover these potentials are present even if temperatures of the thermal bathes are equal (zero affinities) but the system deviates from the thermodynamic limit, e.g mass of gas particles is not negligibly small compared to the math of the system. It is the case of many biological and nanosystems that are in interaction with particles.

According to (\ref{eq:7589}), mass of $\xi$ degree of freedom is rescaled by the factor $G$ (\ref{eq:597965}). The average factor $G$ is a function of linear viscous coefficients (\ref{eq:337587}):
\begin{eqnarray}
\label{eq:17} <G-1>\propto\sum_{\xi'\neq\xi,i}\frac{1}{\rho_{i}\sqrt{mT_{i}}}\gamma_{\xi',i},
\end{eqnarray}
One can hope for experimental verification of this prediction using mass correction measurements\cite{Ekinci2004}. It may be relevant to discrete transition state models\cite{Jarzynski1999}.

According to this work, stability of a thermal ratchet depends on the relative directions of its motion and of its spatial asymmetry. It can be connected to chirality because triangles moving toward the apex and toward the base can not be superimposed on each other (strictly speaking motion along a circular or at least curved path is required)\cite{BARRON1986}. In biology, there is an open question regarding homochirality of the living systems that are composed of L-chiral isomers, though from a chemical point of view R-chiral life would possess the same physical properties\cite{Guijarro2008}. This work provides an example where chirality is broken dynamically\cite{Plasson2007}. It may be interesting for motility based theories for origin of life\cite{Froese2014,Hoffmann2012}.

This work may be useful to estimate the properties of kinetic coefficients of active matter\cite{Bechinger2016} such as asymmetric particles in external flow\cite{Kraft2013,Teeffelen2008}. This work, however, considers interaction with a rear gas, while the experiments are performed in a liquid. In addition, assumption of translation invariant degrees of freedom was used. Both these shortcomings of the approach do not appear explicitly in the results. The results may be extended to  non-linear regime\cite{Gaspard2013}.

Ab initio microscopic approach is essential to predict the steady state of a system with multiple degrees of freedom out of thermal equilibrium. Macroscopic results such as Onsager relations, Prigogine's principle of minimum entropy production at steady state\cite{Prigogine1967}, Onsager Machlup function\cite{Machlup1953} together with recent Jarzynski\cite{Jarzynski1997} and Crooks\cite{Crooks1999} relations for dissipation in a driven system describe the properties of a steady state as a function of the system's symmetries. On the contrary, the microscopic approach makes it possible to calculate the steady state's macroscopic parameters and  symmetries instead of postulating them.

To conclude, even simple dumbbell body out of thermal equilibrium possesses nonintuitive dynamics. Many hope that dynamics of complex mechanical systems may be similar to the behavior of living matter. This work provides the tools to calculate dynamics of a system with arbitrary degrees of freedom. The findings include phenomena such as stability and symmetry breaks imposed by out of equilibrium environment. It is an advance on the path from a single degree of freedom Brownian motors to a multi-degree of freedom Brownian robotics.

\begin{figure}[htb]
\centering
\includegraphics[width=.9\linewidth]{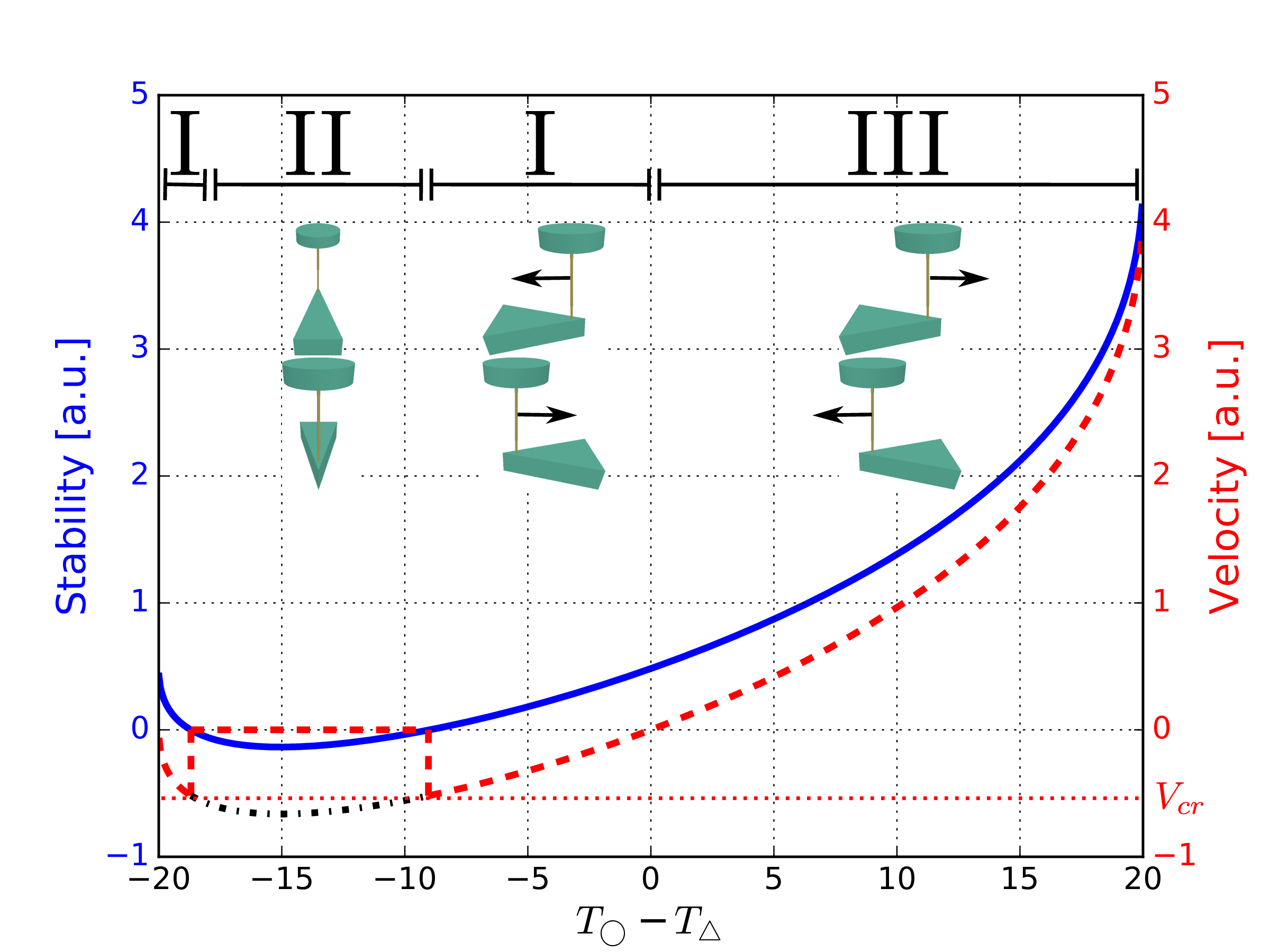}
\caption{\label{fig2}Stability (blue) and velocity (red) of dumbbell structure with asymmetric part composed of an isosceles triangle with apex angle $\theta_{0}=\pi/24$. The rotation axis is located near the apex of the triangle. Velocity (\ref{eq:3785554}) and stability (\ref{eq:6}) presented in arbitrary units ([a.u.]) as the functions of temperature difference $T_{\bigcirc}-T_{\bigtriangleup}$ (under assumption that $T_{\bigcirc}+T_{\bigtriangleup}=20$). There are three regions indicated by roman numerals: First, stable motion toward the base of the triangle. Second, the motion is unstable, while the symmetric orientation of the triangle part is stable $\phi=\pm\pi/2$. The structure which enforces direction along $x$ axis $\phi=0$ or $\phi=\pi$ would move with velocities (dashed black) beyond critical allowed velocity $V_{cr}$. This region may disappear for less acute apex angles and therefore lower velocities. Third, the stable motion toward the apex of the triangle. Stable motion toward the base is surprising because viscous forces in this case act to change the direction of the triangle. It is possible due to potential imposed on the structure by the of thermal equilibrium environment that stabilizes the motion. Nevertheless, motion toward the base is less stable than motion toward the apex. It can be considered as a chiral symmetry break.}
\end{figure}

\begin{figure}
\begin{center}
    \begin{tabular}{c c}
\multicolumn{1}{l}{{\bf\sf A}} & \multicolumn{1}{l}{{\bf\sf B}}\\
\resizebox{0.5\textwidth}{!}{\includegraphics{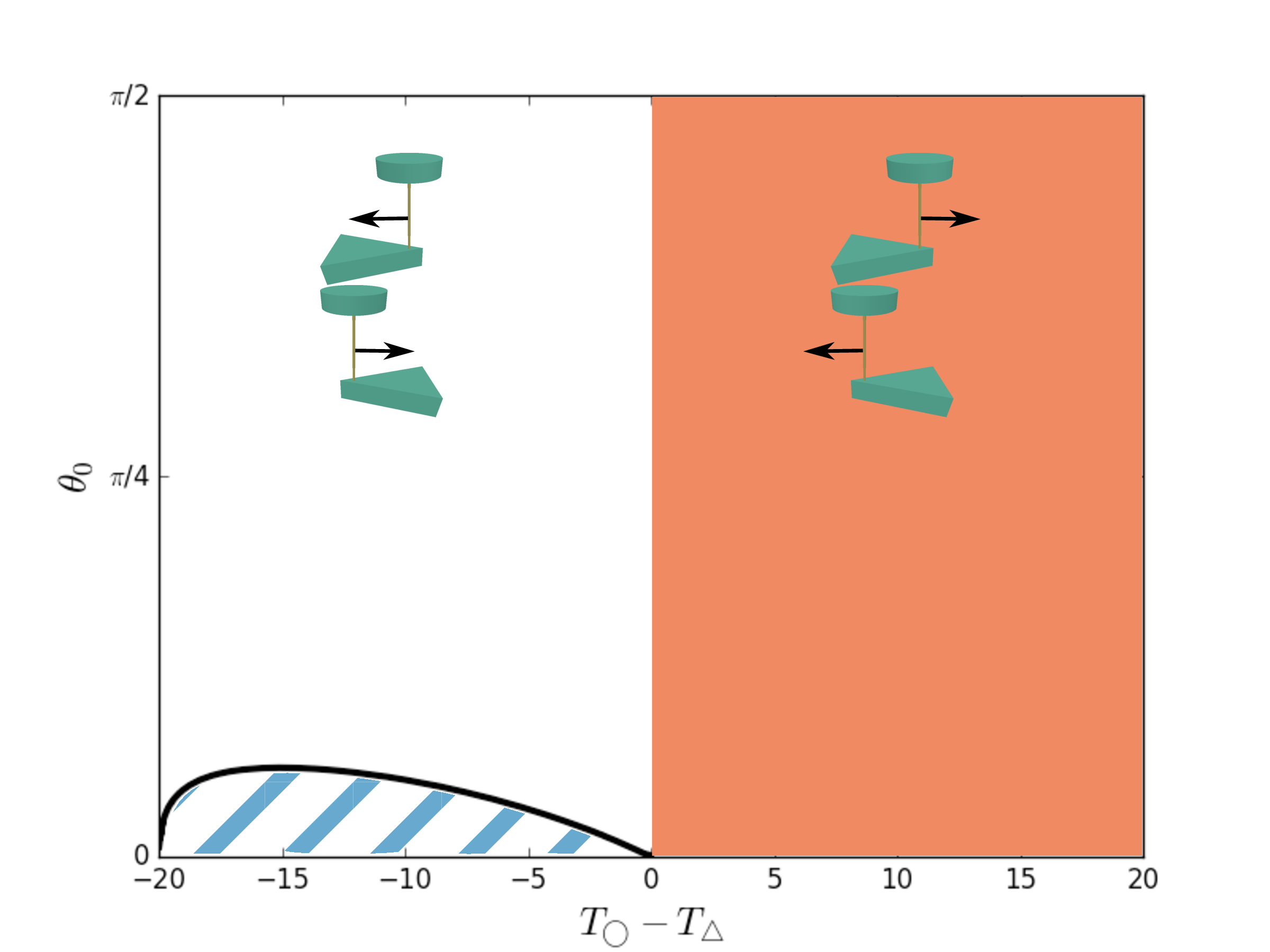}} & \resizebox{0.5\textwidth}{!}{\includegraphics{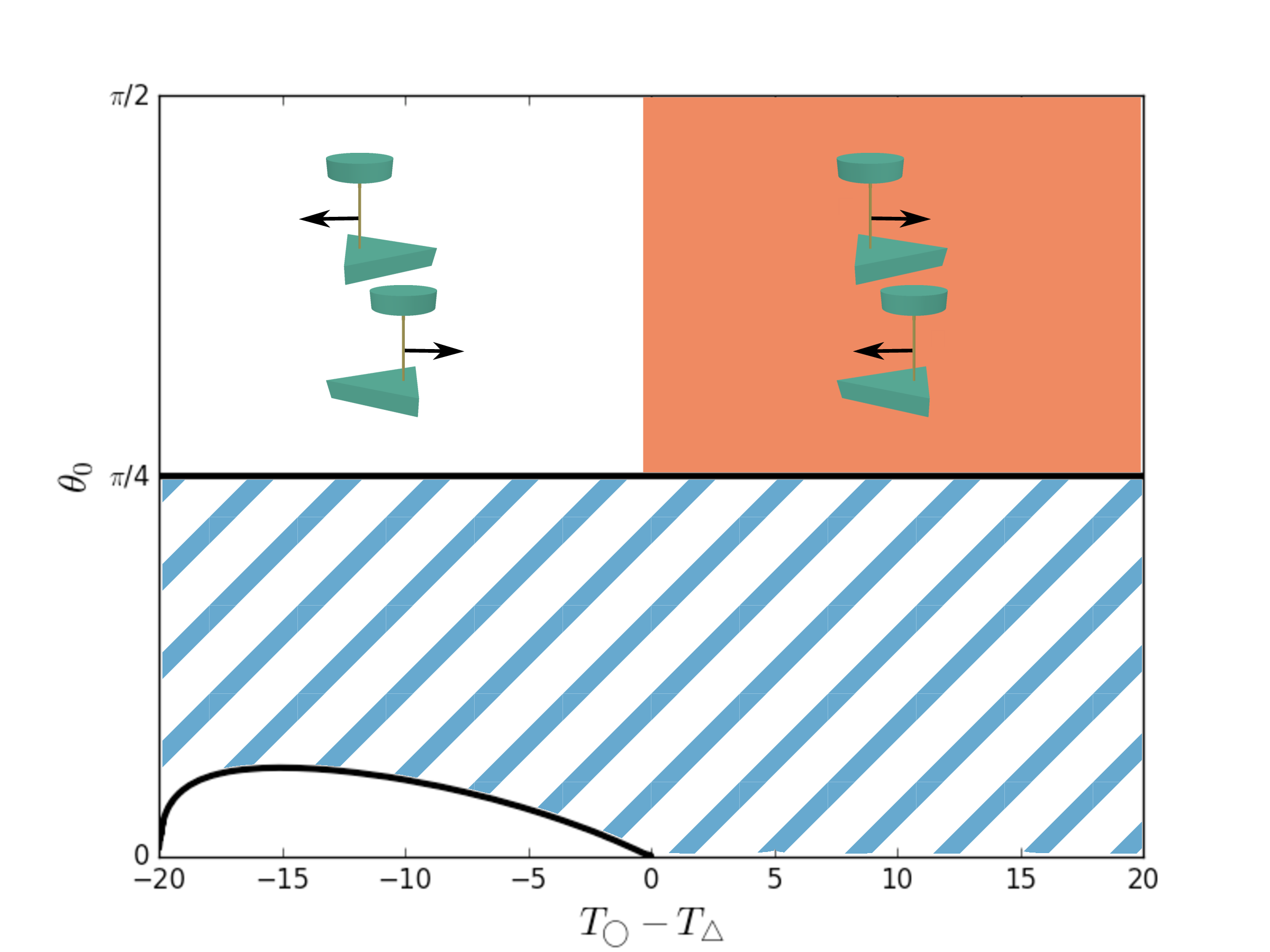}}\\
    \end{tabular}
\caption{Stable orientation and the corresponding translation motion of the dumbbell structure as a function of its shape and the temperatures of the thermal bathes. Asymmetric part is an isosceles triangle with an arbitrary angle of the apex $0<\theta_{0}<\pi$. The phase diagram is presented as a function of $\theta_{0}$ and difference of the temperatures $T_{\bigcirc}-T_{\bigtriangleup}$ (under the assumption that $T_{\bigcirc}+T_{\bigtriangleup}=20$). White region indicates the stable motion toward the base of the triangle. Orange region corresponds to the stable motion toward the apex. No stable motion exists in the dashed region. (A) Rotation axis neat the apex. The motion is stable for the majority of the temperatures and triangle shapes, but for the region of acute apex angle and negative temperature difference (dashed region). This region corresponds to high velocity toward the base of the triangle. (B) Rotation axis near the base. Motion is stable if apex angle is obtuse $2\theta_{0}>\pi/2$. Motion toward the base is also stable in the case of very acute apex angle which corresponds to high velocities.}
\label{fig3}
\end{center}
\end{figure}
\section{Appendix}
\label{sec-1}

To derive transition rate probability $W$ (\ref{eq:7589}) for an arbitrary degree of freedom $X_{\xi}$ and corresponding velocity $\dot{X}_{\xi}$ one should average the possible transitions (\ref{eq:kihkjhk}) over Maxwell gas velocities using as constraints. Transition rate probability $\dot{X}'\rightarrow \dot{X}$ is proportional to the amount of particles that hit the body and amount of possible transitions to specific velocity:
\begin{eqnarray}
  \label{eq:tr1}
  &&W(\dot{X}_{\xi},\dot{X}'_{\xi})=\oint dc \int_{-\infty}^{\infty} dv'_x\int_{-\infty}^{\infty} dv'_yH\left [\left (\dot{X}'_{\xi}-v'\right )e_\perp\right ]\\\nonumber
&&\left |\left (\dot{X}'_{\xi}-v'\right )e_\perp\right | \rho \phi(v'_x,v'_y)\\\nonumber
&&\delta\left [-\Delta\dot{X}_{\xi}-\frac{2\dot{X}_{\xi}-\sum_{\xi'\neq\xi}\frac{\Gamma_{\xi'}(c)}{\Gamma_{\xi}(c)}\dot{X}_{\xi'}-g_{x,\xi}(c)v'_x-g_{y,\xi}(c)v'_y}{1+\frac{M}{m}\left (\frac{G_{\xi}(c)}{\Gamma_{\xi}(c)}\right )^{2}} \right ], 
\end{eqnarray}
where $H$ is Heaviside step function and $\phi$ is Maxwell distribution:
\begin{eqnarray}
  \label{eq:max1}
  \phi(v'_{x},v'_{y})=\frac{m}{2\pi T}\exp\left (\frac{-m(v_{x}^{\prime 2}+v_{y}^{\prime 2})}{2T} \right ),
\end{eqnarray}
From (\ref{eq:42uuy}), (\ref{eq:37768}) and (\ref{eq:kihkjhk}) follows:
\begin{eqnarray}
\label{eq:7}
 (\dot{X}'_{\xi}-v')e_\perp  = -\frac{1}{2}\Delta\dot{X}_{\xi}\Gamma_{\xi}\left (1+\frac{M}{m}\left (\frac{G_{\xi}(c)}{\Gamma_{\xi}(c)}\right )^{2}\right )+\frac{1}{2}\sum_{\xi'\neq\xi}\dot{X}_{\xi'}\Gamma_{\xi'}(c), 
\end{eqnarray}
Using formula:
\begin{eqnarray}
\label{eq:10}
&&\int_{-\infty}^{\infty} dv_{x}\int_{-\infty}^{\infty} dv_{y}\frac{m}{2\pi T}\exp(-\frac{m(v_{x}^{2}+v_{y}^{2})}{2T}\delta(Av_{x}+Bv_{y}+C)=\\\nonumber
&&\sqrt{\frac{m}{2\pi T(A^{2}+B^{2})}}\exp\left (-\frac{m C^{2}}{2T(A^{2}+B^{2})}\right ),
\end{eqnarray}
one gets: 
\begin{eqnarray}
  \label{eq:309837}
  &&W(\dot{X}_{\xi},\Delta \dot{X}_{\xi})=\frac{1}{4}\sum_{i}\rho_{i}\sqrt{\frac{m}{2\pi T_{i}}}\oint dc_{i}\\\nonumber
&&\left | \Delta \dot{X}_{\xi}\right |H\left [\Delta\dot{X}_{\xi}\Gamma_{\xi,i}\left (1+\frac{M_{\xi}}{m}\left (\frac{G_{\xi,i}(c)}{\Gamma_{\xi,i}(c)}\right )^{2}\right )-\sum_{\xi'\neq\xi}\dot{X}_{\xi'}\Gamma_{\xi',i}(c)\right ]\\\nonumber
&&\Gamma_{\xi,i}(c)\frac{1}{\sqrt{g^{2}_{x,\xi,i}+g^{2}_{y,\xi,i}}}\left (1+\frac{M_{\xi}}{m}\left (\frac{G_{\xi}(c)}{\Gamma_{\xi,i}(c)}\right )^{2}\right )^{2}\\\nonumber
&&\exp\left [-\frac{\frac{m}{g^{2}_{x,\xi,i}+g^{2}_{y,\xi,i}}\left ( \dot{X}_{\xi}-\frac{1}{2}\sum_{\xi'\neq \xi}\frac{\Gamma_{\xi',i}}{\Gamma_{\xi,i}}\dot{X}_{\xi'}+\frac{1}{2}\left[ \Delta \dot{X}_{\xi}\left (\frac{M_{\xi} G^{2}_{\xi,i}(c)}{m\Gamma_{\xi,i}^{2}(c)}+1\right )\right ]\right )^{2}}{2T_{i}}\right ]
\end{eqnarray}
where index $i$ goes over all thermal bathes. 

Final expression (\ref{eq:7589}) follows from (\ref{eq:309837}) using (\ref{eq:11}) that rigorously follows from requirement of detailed balance at thermal equilibrium:
\begin{eqnarray}
  \label{eq:8}  P^{eq}(\dot{X}')W(\dot{X},\dot{X}')=P^{eq}(-\dot{X})W(-\dot{X}',-\dot{X}),
\end{eqnarray}
where $P^{eq}(\dot{X})$ is the distribution of velocities of single degree of freedom at equilibrium:
\begin{eqnarray}
  \label{eq:29}
  P^{eq}(\dot{X})\propto\exp\left (-\frac{M\dot{X}^{2}}{2T}\right ),
\end{eqnarray}
in the thermodynamic limit $G\rightarrow 1$.

\bibliographystyle{unsrt}

\end{document}